\documentclass[twocolumn,times]{aastex631}

\usepackage{graphicx}
\usepackage[flushleft]{threeparttable}
\usepackage{blindtext}
\usepackage{amsmath}
\usepackage{mathtools}
\usepackage{multirow}
\usepackage{comment}
\usepackage{natbib} 
\turnoffeditone

\let\oldAA\AA
\renewcommand{\AA}{\text{\normalfont\oldAA}}

\begin{document}
\shortauthors{R. Ouyed}
\def\nar{New Astron.}
\def\na{New Astron.}

\title{Modeling the Light Curve and Spectra of SN 2023aew}

\correspondingauthor{Rachid Ouyed}
\email{rouyed@ucalgary.ca}

\author{Rachid Ouyed}
\affiliation{Department of Physics and Astronomy, University of Calgary, Calgary, AB, Canada}

 \begin{abstract}
We propose that the delayed conversion of a neutron star (NS) into either a quark star (QS) or a hybrid star (HS), occurring approximately $\sim$105-109 days after the supernova (SN) explosion, injects $\sim 2 \times 10^{49}$ erg of thermal energy into the expanded SN ejecta. This energy, delivered over $\sim 40$ days via a quark-nova (QN) shock or spin-down (SpD) power of the HS, can reproduce the photometric and spectral features observed in SN~2023aew.
In this model, the first light-curve peak corresponds to the $^{56}$Ni-powered SN from a stripped-envelope progenitor with a zero-age main sequence mass of $\gtrsim 15$-$16\,M_{\odot}$. The plateau between the two peaks may result from interaction with circumstellar material (CSM), or from SpD power of the NS prior to its conversion. The second peak is powered by the HS, a highly magnetized remnant formed through a quark matter phase capable of sustaining core magnetic fields up to $\sim 10^{18}$\,G.
A scenario involving two phases of SpD power -- first from the NS and later from the HS -- is compelling and supports the hypothesis that some magnetars may be HSs. The SpD energy of the HS powers the QN ejecta -- outer NS layers -- which then transfer energy to the SN ejecta, producing luminous fast blue optical transients (LFBOTs).
This model offers a potential connection between superluminous SNe (SLSNe) and LFBOTs, with implications for high-energy astrophysics, \textit{r}-process nucleosynthesis, and the physics of dense quark matter governed by Quantum Chromodynamics (QCD).
\end{abstract}


\section{Introduction}

In recent years, a number of double-peaked supernovae (SNe) have been discovered, studied, and debated (e.g., \citealt{nakar_2014,nicholl_2016,inserra_2019,galyam_2019,gutierrez_2021,moriya_2024}; see also \citealt{branch_2017} and references therein). One particularly remarkable case is SN 2023aew, detected by the Zwicky Transient Facility (\citealt{bellm_2019,graham_2019,dekany_2020}), which exhibits unusual properties. Initially classified as a Type IIb SN (\citealt{wise_2023}), it underwent a significant re-brightening roughly 90 days after its first peak. During this phase, its spectrum evolved to resemble that of a Type Ic SN (\citealt{frohmaier_2023,hoogendam_2023}). Compared to other double-peaked SNe (\citealt{das_2024}), the interval between the two peaks in SN 2023aew is unusually long. The second peak not only reaches a higher absolute magnitude -- exceeding that of typical luminous Type Ic SNe -- but also exhibits a broader light curve (LC) and slower spectroscopic evolution (\citealt{gomez_2022}), closely resembling SNe Ic around peak and post-peak epochs.

Several powering mechanisms have been proposed for the second peak (see \citealt{sharma_2024,kangas_2024}), including radioactive decay, CSM interaction, magnetar SpD, and delayed fallback accretion onto a central compact object. While CSM interaction faces its own challenges, the magnetar model requires an unusually long delay. Fallback accretion onto a black hole currently appears to be the most favored explanation, although it still demands refinement to fully account for the observed behavior. At present, the true powering mechanism of SN 2023aew’s second peak remains uncertain.

The possibility that NSs may contain deconfined quark matter in their interiors has gained renewed interest, driven by both recent astrophysical observations \citep[e.g.,][and references therein]{zastrow_2025} and theoretical developments in the study of dense QCD matter \citep[e.g.,][]{fraga_2014,gorda_2021}. In particular, QSs -- compact objects composed entirely of deconfined quark matter, possibly in the form of strange quark matter (SQM)—are predicted in scenarios where SQM constitutes the true ground state of hadronic matter \citep{itoh_1970,bodmer_1971,terazawa_1979,witten_1984}. If such objects exist in nature, they would be self-bound and could exhibit an energy per baryon lower than that of $^{56}$Fe.

Alternatively, deconfined quark matter may reside in the cores of HSs, which feature a quark matter core surrounded by a hadronic envelope \citep[e.g.,][]{haensel_1989,glendenning_1997}. The internal structure and stability of HSs are highly sensitive to the equation of state (EOS) of both phases and the microphysical properties of the hadron–quark interface, such as latent heat and the stiffness of each EOS. While the stability of QSs relies on the absolute stability of SQM, HSs may exist within certain parameter regimes even if SQM is metastable \citep{alford_2013,pereira_2018}.

The discovery of pulsars with masses near $2 M_\odot$ \citep{demorest_2010} has imposed stringent constraints on the EOS of dense matter. Nevertheless, quark matter models remain compatible with these mass measurements \citep{alford_2005}. Discriminating between hadronic NSs, HSs, and QSs remain observationally challenging due to degeneracies in macroscopic observables such as the mass–radius relation, cooling behavior, and tidal deformability \citep[e.g.,][]{ozel_2010,weissenborn_2011}. However, recent studies that combine astrophysical constraints with EOS parameterizations continue to support the viability of a phase transition to quark matter in the cores of massive NSs \citep{annala_2023}.

Microphysical simulations of the hadronic-to-quark-matter conversion front reveal the emergence of instabilities that, under certain conditions, may lead to a detonative phase transition \citep{niebergal_2010}. Alternatively, the transition may proceed as a slower, non-detonative conversion, potentially stalling before full deconfinement and resulting in a HS configuration \citep{keranen_2005,ouyed_2018}. Both scenarios can release sufficient energy to eject the outer hadronic layers of the NS (the QN and its ejecta), producing distinct observational signatures such as transient light curves, r-process nucleosynthesis, and bursts of gravitational waves or neutrinos. As such, they offer potential empirical constraints on the properties of QCD matter at densities several times nuclear saturation, where the QCD phase diagram remains poorly understood \citep{guenther_2021}.
These signatures are sensitive to the post-transition structure of the compact object, the nature of the phase conversion, and the time delay between NS formation and its conversion to a HS or QS. 

A delayed transition -- occurring months after core collapse -- of a NS into either a QS or a HS provides a secondary energy source that can re-energize the expanded SN ejecta. This energy, derived from latent heat or rotational energy, has the potential to produce a broad second peak in the LC, with photometric and spectroscopic features distinct from the initial SN peak. This forms the basis of the QN model for double-peaked SNe and superluminous SNe (SLSNe), such as SN 2006gy \citep{leahy_2008}, SN 2009ip \citep{ouyed_2013} and SN 2006oz \citep{ouyed_leahy_2013}; see also \citet{ouyed_2015} for other double-peaked SNe and SLSNe. 

The QN model proposes two main powering mechanisms (see \citealt{ouyed_2022a,ouyed_2022b} for a recent review of the QN's microphysics and macrophysics): (i) \textit{Explosive NS-to-QS conversion}: This scenario involves the complete conversion of the NS into a stable QS, assuming the existence of a stable quark matter state (e.g., \citealt{itoh_1970, bodmer_1971, terazawa_1979, witten_1984}). Each of the  $10^{57}$ baryons in the NS converted releases $\sim 100$ MeV ($\sim 10^{-4}$ erg)  amounting to $\sim 10^{53}$ erg as latent heat.  A fraction  is converted into kinetic energy ($E_{\rm QN, KE}\sim 10^{52}$~erg), which drives the ejection of the NS crust (the QN ejecta, with mass $M_{\rm QN, ej} \sim 10^{-5}\,M_\odot$). A portion of this kinetic energy is thermalized within the SN envelope, resulting in a re-brightening of the LC; (ii) \textit{Non-explosive NS-to-HS conversion}: In this scenario, only the core of the NS is converted, leading to the formation of a highly magnetized HS. During this partial conversion, approximately $M_{\rm QN, ej}\sim 0.05\,M_\odot$ of the outer NS layers are ejected with a kinetic energy of $E_{\rm QN, KE}\sim 5 \times 10^{50}$~erg \citep{keranen_2005}. Here, the energy powering the SN ejecta comes primarily from the SpD of the resulting magnetar-strength HS. For this mechanism to be viable, a means of terminating the SpD emission is required (e.g., collapse into a black hole or leakage of energy via hard emission). When either the QN shock or HS SpD energy is deposited into the already-expanded SN ejecta after a delay of $\sim 105$-$109$ days, the resulting thermal input -- on the order of $\sim 2\times 10^{49}$~erg over $\sim 40$ days -- can naturally reproduce the LC and spectral features observed in SN~2023aew.

The paper is organized as follows: In \S \ref{sec:lc}, We model the LC of SN 2023aew by considering an SESN (yielding the first peak) followed by the delayed heat deposition from the conversion of the NS responsible for the second peak. The observed plateau between the two peaks, along with the post-peak bumps, can be reproduced by including either SN-CSM interaction or NS SpD prior to conversion with the latter favored. The spectral properties are computed in \S \ref{sec:spectrum} near peak 1 and peak 2 which captures key
 properties of the observed one. A discussion and a conclusion are given in  \S \ref{sec:discussion}.

\section{The light curve}
\label{sec:lc}

The first peak is powered by $^{56}$Ni decay with an ejecta of mass $M_{\rm ej, 1}$ moving at a speed $v_{\rm ej, 1}$.
   The diffusion timescale is $t_{\rm d, 1}= \sqrt{2\kappa_{\rm ej, 1} M_{\rm ej, 1}/\beta c v_{\rm ej, 1}}$ 
with $\beta=4\pi^3/9$  a geometric correction factor (Arnett 1982) and $c$ the speed of light. Hereafter, subscript 1 (2) refers to quantities at peak 1 (peak 2). The optical opacity we take to be $\kappa_{\rm ej, 1} = 0.1$ cm$^2$ g$^{-1}$ representative of  stripped-envelope SN (SESN; e.g., \citealt{wheeler_2015}).  The progenitor's initial radius we set to $R_{0, 1} = 140 R_{\odot}$ noting that the results presented in this work are insensitive to the exact value as long as $R_0 < 10^3R_{\odot}$.  The gamma-ray optical depth is $A_\gamma =30$  with practically most gamma rays and positrons trapped (e.g. Eq.(9) in \citealt{chatzopoulos_2012}).  

\begin{figure}[t!]
\begin{center}
\includegraphics[scale=0.35]{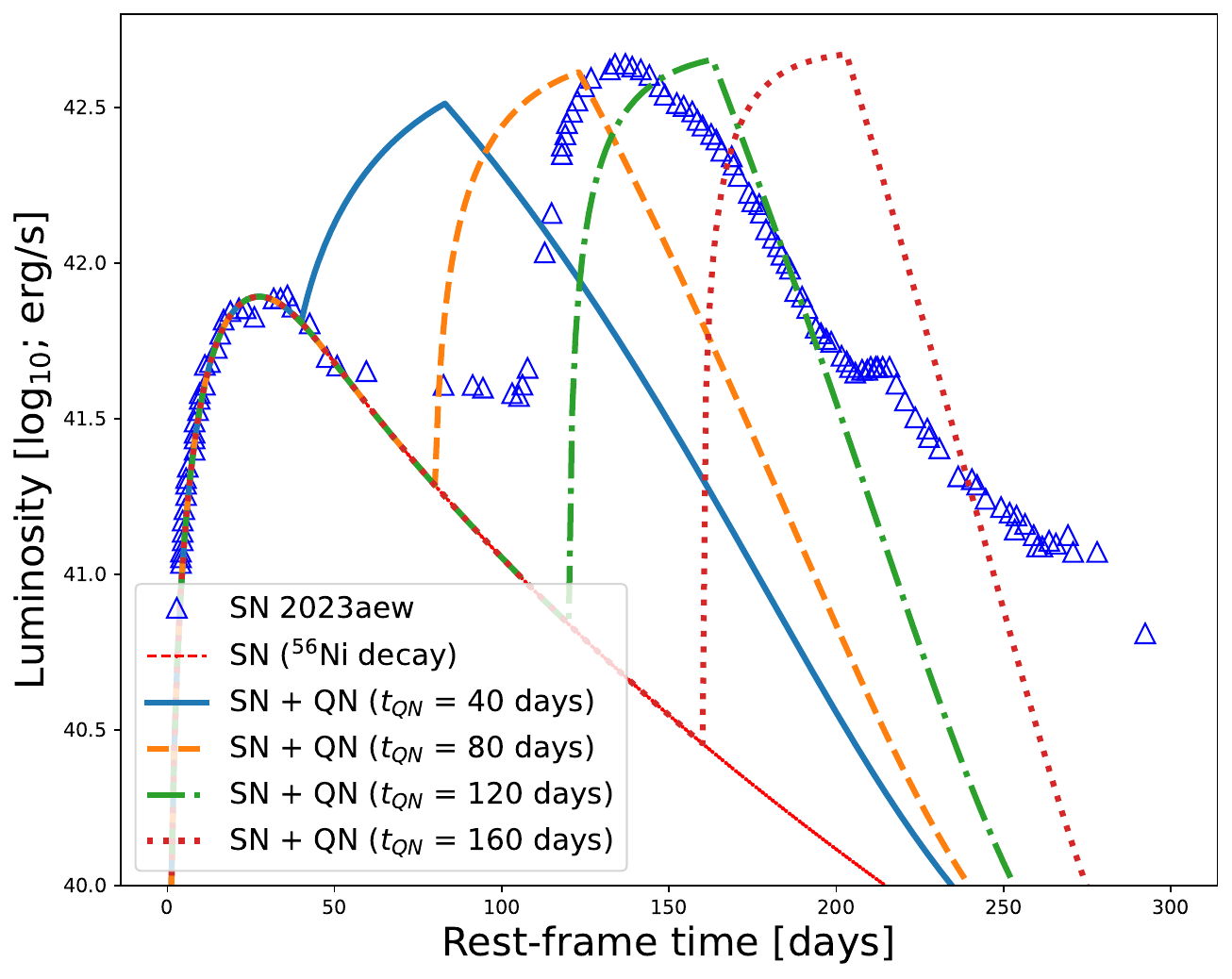}
\includegraphics[scale=0.35]{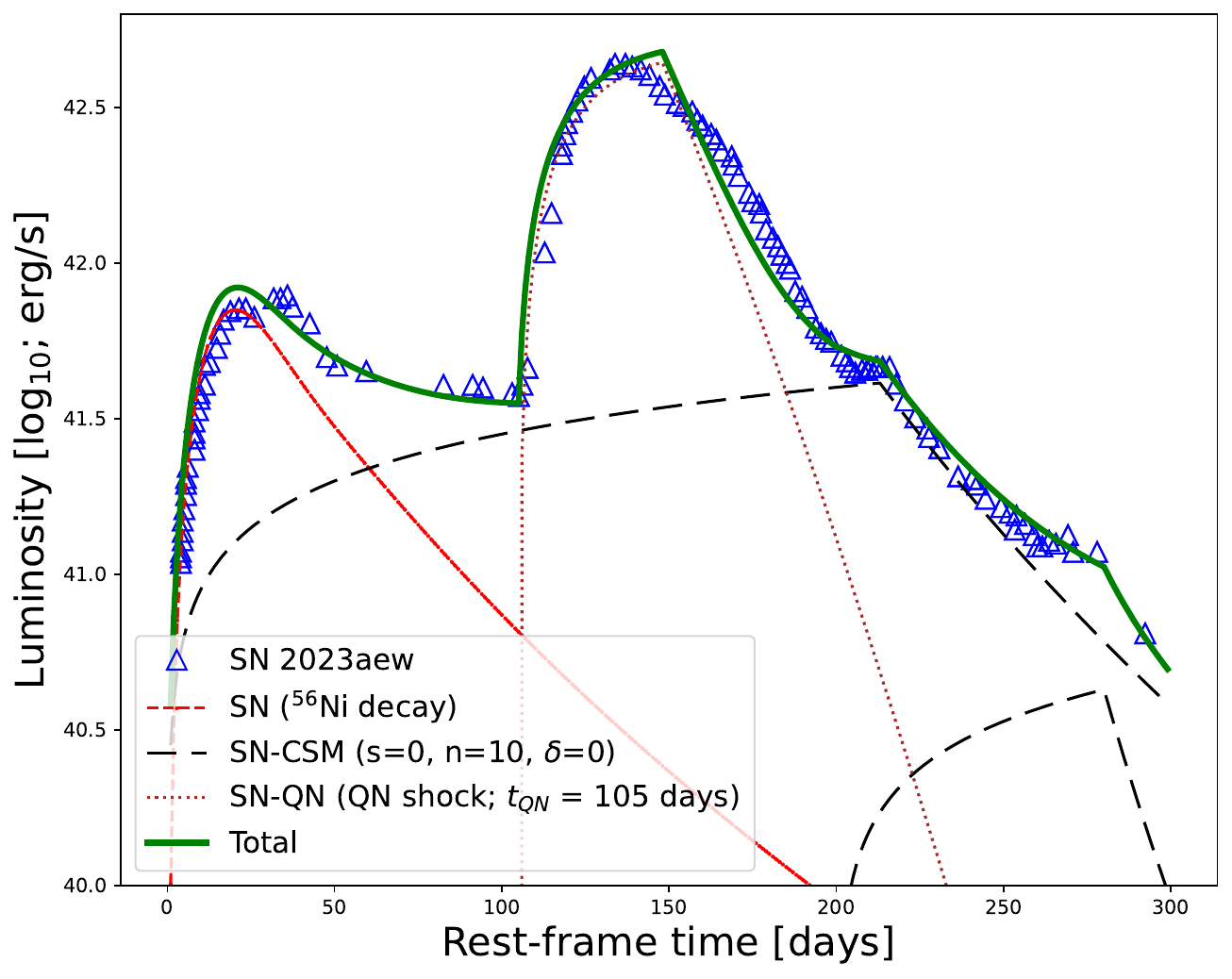}
\caption{{\bf Top panel}: The pseudo-bolometric LC from our model. The first peak is powered by $^{56}$Ni, while the second peak arises from re-shocked SN ejecta due to delayed energy injection following the conversion of the NS (see \S \ref{sec:lc}). The plot shows the re-brightening of the SN ejecta (second peak) for increasing time delays, $t_{\rm QN}$, between the SN and the QN. The pseudo-LC of SN 2023aew (without error bars) is shown for reference (\citealt{sharma_2024,kangas_2024}).
{\bf Bottom panel}: Fit to the LC of SN 2023aew with $t_{\rm QN} = 105$. The plateau and the bumps beyond the second peak are due to the interaction between the SN ejecta and the CSM, starting at 
$t=0$ days for the first shell and at $t\sim 200$ days for the second shell (see Table \ref{table:table1} for relevant parameters).}
\label{fig:LC-QN}
\end{center}
\end{figure}

The resulting fit to the pseudo-bolometric LC is shown in the top panel in Figure \ref{fig:LC-QN}. It is 
  obtained with an ejecta mass of $M_{\rm ej, 1}=5.5M_{\odot}$ moving at $v_{\rm ej, 1}=10^4$ km s$^{-1}$ with  $0.06M_{\odot}$ of $^{56}$Ni (see Table \ref{table:table1}).  The corresponding SN energy, $\sim 4\times 10^{51}$ erg, suggests an energetic SN.  In this  paper, the parameters represent a good manually-obtained fit rather than best fit from minimizing $\chi^2$.  The pseudo-bolometric LC we fit is derived from multi-band photometry, with uncertainties inflated through error propagation, filter interpolation, and assumptions about the spectral energy distribution (\citealt{sharma_2024,kangas_2024}). A rigorous $\chi^2$ treatment would involve direct fitting to the multi-band photometry, preserving the reported observational uncertainties -- an approach that lies beyond the scope of this work.

 %
 %
\begin{table*}[t!]
  \centering
\caption{SN 2023aew lightcurve fitting: QN shock}
 \centering
 { 
\footnotesize
  \begin{tabular}{|c|c|c|c||c|c|c||c|c|c|c|c|}\hline
  \multicolumn{4}{|c||}{SN parameters} &  \multicolumn{3}{|c||}{QN  parameters}  &  \multicolumn{5}{|c|}{CSM  parameters} \\\hline
  $M_{\rm ej, 1}(M_{\odot})$ &   $v_{\rm ej, 1}$ (km s$^{-1}$) & $M_{\rm Ni}(M_{\odot})$ & $\kappa_{\rm ej, 1}$ (cm$^{2}$ g$^{-1}$)  &
  $t_{\rm QN}$ (days) & $E_{\rm sh}$ (erg) & $t_{\rm sh}$ (days)  & 
  $s$ & $n$ & $\delta$ & $\zeta_{\rm CSM}$ & $t_{\rm t}$ (days)\\\hline
    \multicolumn{12}{|c|}{No SN-CSM interaction (top panel in Figure \ref{fig:LC-QN})} \\\hline
   5.5 &  10,000 & 0.06 &  0.1 &  40,80,120,160& $2\times 10^{49}$ & $40$ &   -- & -- & -- &  --- & --- \\\hline
    \multicolumn{12}{|c|}{With SN-CSM interaction (bottom panel in Figure \ref{fig:LC-QN})} \\\hline
    4.0 & 12,000& 0.04 &  0.08 &   105& $1.8\times 10^{49}$ & 43 &  0 & 10 & 0&  $9\times 10^{-3}$ & 213\\\hline
   4.0 & 12,000& 0.04 &  0.08 &   105& $1.8\times 10^{49}$ & 43 &   0 & 10 & 0&  $3.5\times 10^{-4}$ & 280\\\hline
  \end{tabular}
  }
  \label{table:table1}
  \end{table*}
 %
 %
 %
 %
  \begin{table*}[t!]
  \centering
\caption{SN 2023aew lightcurve fitting: HS spin-down power $+$ SN-CSM interaction.}
 { 
\footnotesize
  \begin{tabular}{|c|c|c|c||c|c|c|c||c|c|}\hline
  \multicolumn{4}{|c||}{SN parameters}&  \multicolumn{4}{|c||}{HS  parameters}   & \multicolumn{2}{|c|}{CSM parameters$^{\dagger}$}    \\\hline
    \multicolumn{10}{|c|}{Without HS undulations  (top panel in Figure \ref{fig:LC-CSM-HS})} \\\hline
  $M_{\rm ej, 1}(M_{\odot})$ &   $v_{\rm ej, 1}$ (km s$^{-1}$) & $M_{\rm Ni}(M_{\odot})$ & $\kappa_{\rm ej, 1}$ (cm$^{2}$ g$^{-1}$)   &
  $t_{\rm HS}$ (days) & $P_{\rm HS}$ (ms) & $B_{\rm HS}$ (G)  &  $A_{\rm HS}$ (s$^2$) & 
   $\zeta_{\rm CSM}$ & $t_{\rm t}$ (days) \\\hline
      5.0 & 12,000& 0.04 &  0.08 &   109& 31 & $8\times 10^{14}$ & $2\times 10^{13}$ &  $7\times 10^{-3}$ & 213 \\\hline
       \multicolumn{10}{|c|}{With HS undulations  (bottom panel in Figure \ref{fig:LC-CSM-HS})} \\\hline
  $M_{\rm ej, 1}(M_{\odot})$ &   $v_{\rm ej, 1}$ (km s$^{-1}$) & $M_{\rm Ni}(M_{\odot})$ & $\kappa_{\rm ej, 1}$ (cm$^{2}$ g$^{-1}$)   &
  $t_{\rm HS}$ (days) & $P_{\rm HS}$ (ms) & $B_{\rm HS}$ (G)  &  $A_{\rm HS}$ (s$^2$) & 
   $\zeta_{\rm CSM}$ & $t_{\rm t}$ (days) \\\hline
      5.0 & 12,000& 0.043 &  0.08 &   109& 28 & $6\times 10^{14}$ & $2\times 10^{13}$ &  $4\times 10^{-3}$ & 150 \\\hline
  \end{tabular}
  }\\
  $^{\dagger}$ Here, $s=0, n=10$ and $\delta=0$ for the CSM. 
   \label{table:table2}
   \end{table*}
  %
 %
\begin{figure}[t!]
\begin{center}
\includegraphics[scale=0.35]{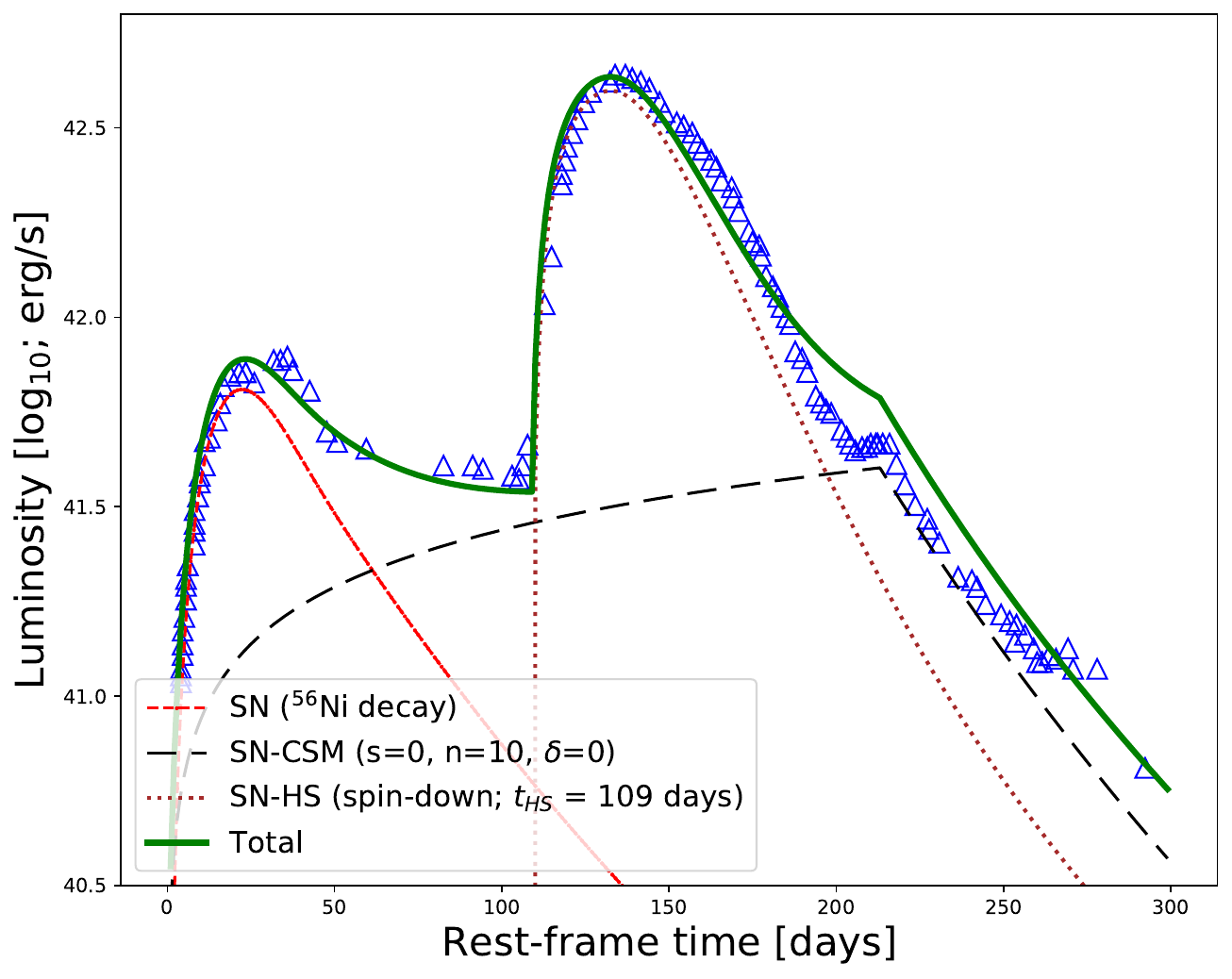}
\includegraphics[scale=0.35]{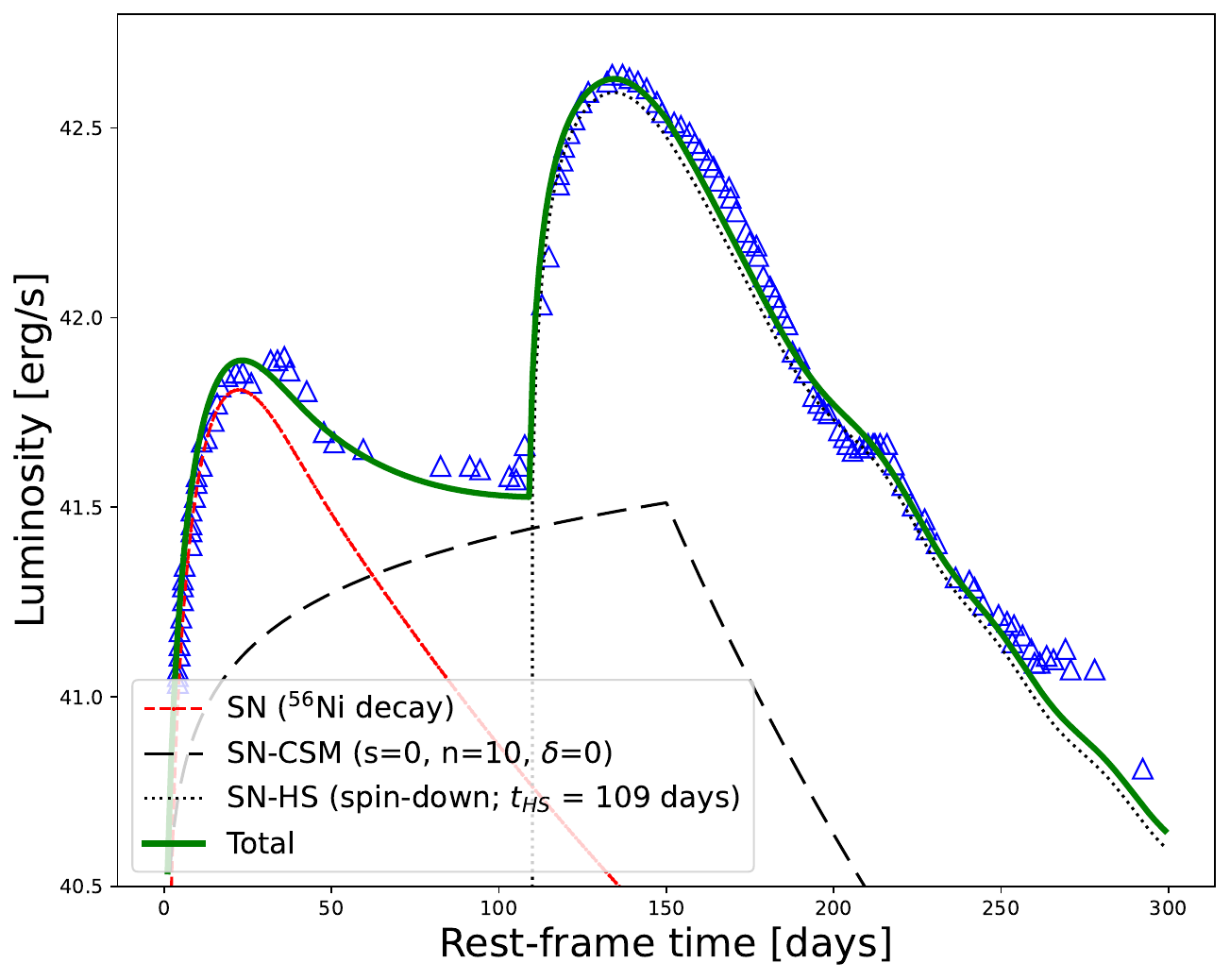}
\caption{Fit to the pseudo-bolometric luminosity of SN 2023aew using a combined  $^{56}$Ni, SN-CSM interaction and HS spin-down power ($t_{\rm HS} = 109$ days).
{\bf Top panel}: Model includes hard emission leakage from the HS (see \S \ref{sec:lc}). {\bf Bottom panel}: Model includes hard emission leakage from a precessing  HS. See Table \ref{table:table2}  for parameter values.}
\label{fig:LC-CSM-HS}
\end{center}
\end{figure}
%
 %
 \subsection{The QN shock}
 \label{sec:QN-shock}
 
 To model the second peak, we consider a SESN progenitor with a zero-age main-sequence (ZAMS) mass $M_{\rm prog.}$, which underwent mass ejection episodes prior to the SN, resulting in the formation of a CSM. The explosion left behind a slowly rotating, low-mass NS (here, $M_{\rm NS}=1.5M_{\odot}$), with typical values for the spin period ($P_{\rm NS} > 10$ ms) and surface magnetic field ($B_{\rm NS}< 10^{14}$ G) such that its contribution to the SN bump is negligible.

We assume that by the time the QN occurs, the SN ejecta has already mixed with some of the material expelled during the pre-SN mass-loss phase, while emission from the SN-CSM interaction is neglected for now. The mixed ejecta of mass $M_{\rm ej, 2}$ is moving at a speed $v_{\rm ej, 2}= v_{\rm ej, 1}\times M_{\rm ej, 1}/M_{\rm ej, 2}$. 
The diffusion timescale for the mixed ejecta is $t_{\rm d, 2}= t_{\rm d, 1}\times (M_{\rm ej, 2}/M_{\rm ej, 1})\times (\kappa_{\rm ej, 2}/\kappa_{\rm ej, 1})^{1/2}$ where $\kappa_{\rm ej, 2}$ is the opacity of the mixed ejecta.
The SN ejecta's size by then is $R_{0, 2}= R_0 + v_{\rm ej, 2}t_{\rm QN}$. In both scenarios considered in this section, relativistic propagation allows the QN ejecta or the HS wind to catch up with the SN ejecta on relatively short timescales $R_{0, 2}/c << t_{\rm QN}$; i.e. $v_{\rm ej, 2}/c<<1$.

An amount of heat $E_{\rm sh}$ is injected (here via the QN shock)  into the expanded ejecta starting at $t_{\rm QN}$ after the SN and at a constant luminosity $L_{\rm sh}$ over a time $t_{\rm sh}$.  
This gives a LC rising to a peak luminosity $L_{\rm peak, 2}=E_{\rm sh}/t_{\rm sh}$ and declines thereafter on the characteristic diffusion timescale
$t_{\rm d, 2}$ (see Eq. (7) in \citealt{chatzopoulos_2012}).   For modeling the re-brightening, SN 2023aew serves as a reference for the height and width of the second peak.
The top panel in Figure 1 shows the resulting second bumps for $E_{\rm sh}=2\times 10^{49}$ erg, $t_{\rm sh}=40$ days and different time delays $t_{\rm QN}$.  Shown is the case of $M_{\rm ej, 2}=2  M_{\rm ej, 1}$  and $\kappa_{\rm ej, 2}=3\kappa_{\rm ej, 1}$ so that  $t_{\rm d, 2}\sim 3.7 t_{\rm d, 1}$  (see Table \ref{table:table1} for the parameters used).    

 The fit to the pseudo-bolometric LC of SN 2023aew is shown in the bottom panel in Figure \ref{fig:LC-QN} with relevant parameters listed in Table \ref{table:table1};
 the time delay is $t_{\rm QN}= 105$ days. The  plateau between the two peaks and the bumps beyond the second peak appear when taking into account emission from the
SN-CSM interaction. We use the analytical LC model of \citet{moriya_2013} which assumes a constant CSM velocity $v_{\rm w}$ and a CSM density profile $\rho_{\rm CSM} = D r^{-s}$ where $D$ is a constant. We take $s=0$ (i.e. a CSM shell with a flat density profile) so that  $D= \rho_{\rm csm}$.  
 The SN  ejecta is defined by its kinetic energy $E_{\rm ej, 1}$ and its mass $M_{\rm ej, 1}$. Its has a double power-law profile for the density of homologously expanding ejecta ($\rho_{\rm ej, 1}\propto r^{-n}$ outside and $\rho_{\rm ej, 1}\propto r^{-\delta}$ inside).  Another parameter in these models is the conversion efficiency from kinetic energy to radiation, $\zeta_{\rm CSM}$. 
 We adopt an outer density slope $n=10$  and an inner density slope $\delta =0$ (e.g. \citealt{matzner_1999}).
 
For the fit to the SN 2023aew LC, we first consider a single CSM shell that begins to interact with the SN ejecta at $t = 0$.
 With $s=0$ and for $n=10$ and $\delta=0$, the resulting luminosity (ignoring diffusion) increases over time as $t^{1/2}$  as expected until time $t_{\rm t}$
which is the time when the interacting region reaches down to the inner ejecta (see Eq. 8 in \citet{moriya_2013}).  
When fitting SN 2023aew, we take $t_{\rm t}$ to correspond to the point marking the edge (or drop) of the post-peak 2 bump (i.e. $\sim 213$ days) which gives  $D=\rho_{\rm csm}\sim 10^{-16}$ g cm$^{-3}$. The mass of the shell,  $M_{\rm shell, 1}\sim \frac{4\pi}{3} \rho_{\rm csm} (v_{\rm ej, 1} t_{\rm t})^3$, is of the order of a few solar masses  consistent with the $M_{\rm ej, 2}=2M_{\rm ej, 1}$ assumption we used.    As shown in Table 2, including the SN-CSM interaction meant reducing the SN ejecta mass and $^{56}$Ni content. The interaction with a second shell, beginning around $t\sim 200$ days after the SN explosion, provides a good fit to both the second bump and the subsequent decline following the second peak.
  The $t_{\rm t}=280$ days for shell 2  gave $\rho_{\rm csm}\sim 10^{-15}$ g cm$^{-3}$ or a shell mass $M_{\rm shell, 2}\sim \frac{4\pi}{3} \rho_{\rm csm} (v_{\rm ej, 2} t_{\rm t})^3 \sim 6M_{\odot}$.  This suggests that the progenitor had no less than two mass-loss episodes prior to the SN event and it had a ZAMS mass of at least $M_{\rm prog.}= M_{\rm ej, 2}+ M_{\rm shell, 2}+M_{\rm NS}\sim$ 15-16$M_{\odot}$. The SN-CSM interaction requires a very low radiation conversion efficiency, $\zeta_{\rm CSM}$, particularly in the case of the second shell (see, however, \S \ref{sec:HS-spindown}).
 
 In scenario (i), re-heating occurs when the relativistic QN ejecta catches up to and impacts the mixed ejecta, generating a QN shock.
 In this case, the conversion efficiency from kinetic energy (of the QN ejecta) to heat is very low $\zeta_{\rm QN}=E_{\rm sh}/E_{\rm QN, KE}\sim 10^{-3}$
 but expected given the very low envelope density (e.g. \citealt{ouyed_2013,ouyed_2015}).
 Also,  $t_{\rm sh}\sim v_{\rm ej, 2}t_{\rm QN}/v_{\rm QN, sh}\sim 43$ days if  the QN shock velocity is $v_{\rm QN, sh}\sim 2 v_{\rm ej, 2}\sim v_{\rm ej, 1}$  which is comparable to $(2 E_{\rm QN, KE}/M_{\rm ej, 2})^{1/2}\sim 1.2\times 10^4$ km s$^{-1}$.

   \subsection{The HS spin-down (SpD) power}
 \label{sec:HS-spindown}
 
 If the neutron-to-quark transition proceeds as a smooth crossover, no latent heat is released, leaving the HS scenario as the alternative. In this scenario (ii), re-heating of the expanded SN ejecta is powered by SpD energy, with the HS inheriting the NS's spin period and acquiring a strong magnetic field (see \S \ref{sec:discussion}).
The HS period at $t_{\rm HS}$ is  $P_{\rm HS}= P_{\rm NS} \times (1+t_{\rm HS}/t_{\rm NS, SpD})^{1/3}$ (e.g. \citealt{michel_1970}) with $t_{\rm NS, SpD}\simeq 5.8\ {\rm yrs}\times (P_{\rm NS}/30\ {\rm ms})^2 (10^{14}\ {\rm G}/B_{\rm NS})^2$ the parent NS SpD timescale.
While shock heating in scenario (i) is finite and determined by the size of the ejecta, scenario (ii) requires a mechanism to shut off the SpD energy within the SpD timescale, 
$t_{\rm HS, SpD}$  (e.g., if the HS collapses into a black hole or via interaction with the QN ejecta). Without such a mechanism, the post-peak LC decline is difficult to reproduce, even when accounting for significant hard emission leakage (here a leakage parameter $A_{\rm HS}=2\times 10^{13}$ s$^2<< 3 \kappa_{\rm ej, 2} M_{\rm ej, 2}/4\pi v_{\rm ej, 2}^2$; e.g. \citealt{wang_2015}). See top panel in Figure \ref{fig:LC-CSM-HS} with the corresponding parameters listed in Table \ref{table:table2}. 

  The HS case offers another plausible explanation of the late bumps  if we consider that the strong core magnetic field distorts the HS and induces an ellipticity $e_{\rm HS}\sim 10^{-8}$ (e.g. \citealt{haskell_2008} and references therein) with a corresponding precession period $P_{\rm HS}^{\rm prec.}\sim P_{\rm HS}/e_{\rm HS}$.
   The late undulations in the bottom panel in Figure \ref{fig:LC-CSM-HS} are obtained following a simple Gaussian prescription where the input SpD luminosity of the non-precessing HS is modulated as $L_{\rm HS, SpD}^{\rm prec.}(t)= L_{\rm HS, SpD}(t)\times \left(1+ \alpha_{\rm HS} \exp{[- 0.5 (\frac{t- P_{\rm HS}^{\rm prec.}}{\beta_{\rm HS} P_{\rm HS}^{\rm prec.}})^2}]\right)$.
     The plateau is still due to the SN-CSM interaction with $t_{\rm t}=150$ days so that $M_{\rm ej, 2}=2M_{\rm ej, 1}$ and $\kappa_{\rm ej, 2}=3\kappa_{\rm ej, 1}$ remains valid.
   Hydrodynamic simulations may be necessary to capture effects that are overlooked in the more simplified treatment of the SN-CSM+HS (with undulations) scenario.
     
  %
 %
  \begin{table*}[t!]
  \centering
\caption{SN 2023aew lightcurve fitting: HS spin-down power (with undulations) $+$ SN-NS interaction.}
 { 
\footnotesize
  \begin{tabular}{|c|c|c|c||c|c|c|c|c||c|}\hline
  \multicolumn{4}{|c||}{SN parameters$^\dagger$}&  \multicolumn{5}{|c||}{HS  parameters$^{\dagger\dagger}$}   & \multicolumn{1}{|c|}{NS parameters$^{\dagger\dagger\dagger}$}    \\\hline
    \multicolumn{10}{|c|}{Top panel in Figure \ref{fig:LC-NS-HS}} \\\hline
  $M_{\rm ej}(M_{\odot})$ &   $v_{\rm ej}$ (km s$^{-1}$) & $M_{\rm Ni}(M_{\odot})$ & $\kappa_{\rm ej}$ (cm$^{2}$ g$^{-1}$)   &
  $t_{\rm HS}$ (days) & $B_{\rm HS}$ (G)  &  $A_{\rm HS}$ (s$^2$) &  $\alpha_{\rm HS}$ &  $\beta_{\rm HS}$ & 
  $P_{\rm NS}$ (ms)   \\\hline
      15.0 & 12,000& 0.03 &  0.02 &   108&  $4.2\times 10^{14}$ & $1.8\times 10^{13}$ & 0.4 & 0.3 & 
       22.3   \\\hline
    \multicolumn{10}{|c|}{Bottom panel in Figure \ref{fig:LC-NS-HS}} \\\hline
  $M_{\rm ej}(M_{\odot})$ &   $v_{\rm ej}$ (km s$^{-1}$) & $M_{\rm Ni}(M_{\odot})$ & $\kappa_{\rm ej}$ (cm$^{2}$ g$^{-1}$)   &
  $t_{\rm HS}$ (days) &  $B_{\rm HS}$ (G)  &  $A_{\rm HS}$ (s$^2$) & $\alpha_{\rm HS}$ &  $\beta_{\rm HS}$ & 
  $P_{\rm NS}$ (ms)    \\\hline
      18.0 & 12,000& 0.03 &  0.02 &   108& $2.4\times 10^{14}$ & $1.1\times 10^{13}$ & 0.3 & 0.3 & 
     16.8  \\\hline
  \end{tabular}
  }\\
  $^\dagger$  There is no CSM interaction or sweeping involved, so only a single ejecta component is considered.\\
  $^{\dagger\dagger}$ $P_{\rm HS}= P_{\rm NS} \times (1+t_{\rm HS}/t_{\rm NS, SpD})^{1/3}$ (e.g. \citealt{michel_1970}) with $t_{\rm NS, SpD}\simeq 5.8\ {\rm yrs}\times (P_{\rm NS}/30\ {\rm ms})^2 (10^{14}\ {\rm G}/B_{\rm NS})^2$.\\
$^{\dagger\dagger\dagger}$ $B_{\rm NS}=7.5\times 10^{13}\ {\rm G}~ (P_{\rm NS}/20\ {\rm ms})^2$.   The NS hard emission is fully trapped.
   \label{table:table3}
   \end{table*}
    %
 %
 
 %
 %
\begin{figure}[t!]
\begin{center}
\includegraphics[scale=0.35]{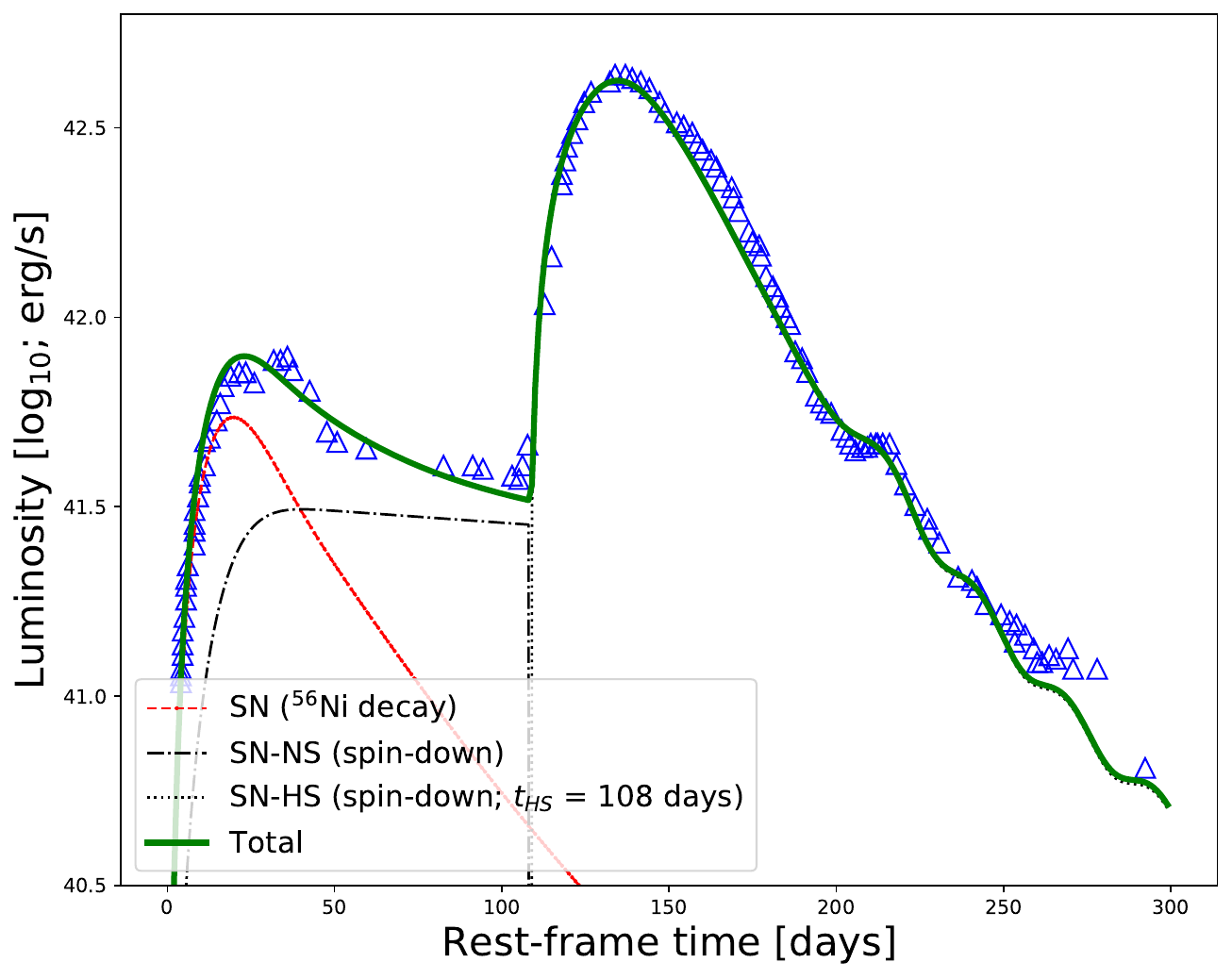}
\includegraphics[scale=0.35]{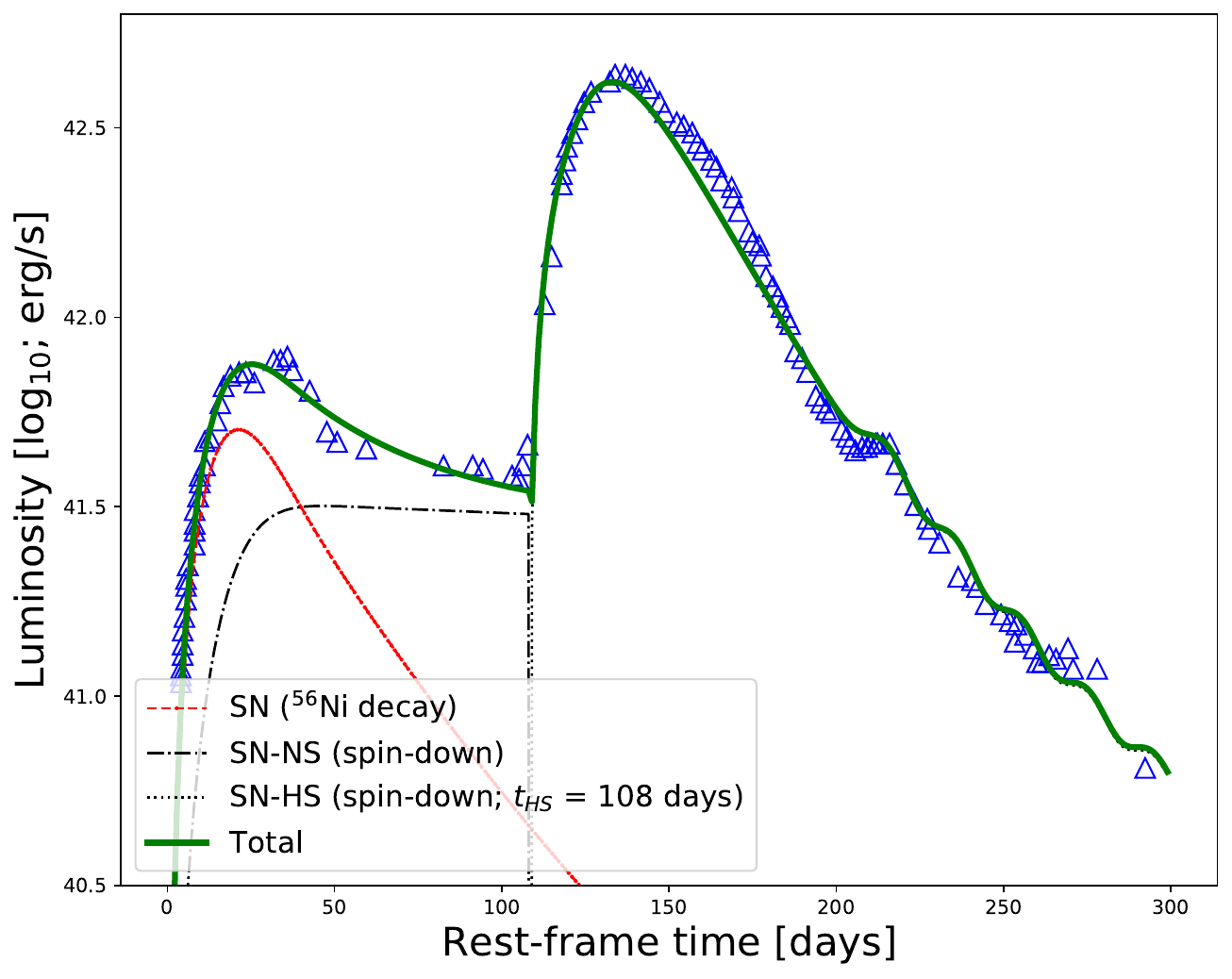}
\caption{Fit to the pseudo-bolometric luminosity of SN 2023aew using spin-down power and a precessing HS with $t_{\rm HS} = 109$ days.
There is no CSM interaction or sweeping involved. The plateau is fit using NS spin-down power
with $P_{\rm NS}= 22.3$ s ({\bf top panel}) and $P_{\rm NS}= 16.8$ s ({\bf bottom panel}). See Table \ref{table:table2}  for parameter values.}
\label{fig:LC-NS-HS}
\end{center}
\end{figure}
%
 %
  %
 %
  \begin{table*}[t!]
  \centering
\caption{SN 2023aew lightcurve fitting: Reprocessed LFBOT $+$ ejecta collision
(bottom panel in Figure \ref{fig:LC-LFBOTS}).}
 { 
\footnotesize
  \begin{tabular}{|c||c|c|c|c||c|c|c|}\hline
    \multicolumn{1}{|c||}{NS  parameters} & 
   \multicolumn{4}{|c||}{LFBOT  parameters$^{\dagger\dagger}$} &
    \multicolumn{3}{|c|}{Ejecta collision (late bump) parameters} \\\hline
   
   $P_{\rm NS}^{\dagger}$ (ms) &  $B_{\rm HS}$ (G)  &    $E_{\rm QN, KE}$ (erg) &  $M_{\rm QN, ej} (M_{\odot})$ & $\kappa_{\rm QN, ej}$ (cm$^2$ g$^{-1}$) &  $E_{\rm shock}$  & $t_{\rm shock}$ (days) & $\kappa$ (cm$^2$ g$^{-1}$)\\\hline
      3.5 &   $5.5\times 10^{14}$  & $6.5\times 10^{50}$ & 0.12 & 0.2   &  $4\times 10^{-3}E_{\rm QN, KE}$ & 33 & 0.3 \\\hline
    \end{tabular}
  }\\
  $^{\dagger}$$B_{\rm NS}=7.5\times 10^{13}\ {\rm G}~ (P_{\rm NS}/20\ {\rm ms})^2$.   $A_{\rm NS}= 3\kappa_{\rm ej}M_{\rm ej}/4\pi v_{\rm ej}^2$; 
  $^{\dagger\dagger}$$P_{\rm HS}= P_{\rm NS} \times (1+t_{\rm HS}/t_{\rm NS, SpD})^{1/3}$. $A_{\rm HS}= 3\kappa_{\rm QN, ej}M_{\rm QN, ej}/4\pi v_{\rm QN, ej}^2$.\\
   \label{table:table4}
   \end{table*}
    %
 %

  %
 %
\begin{figure}[t!]
\begin{center}
\includegraphics[scale=0.35]{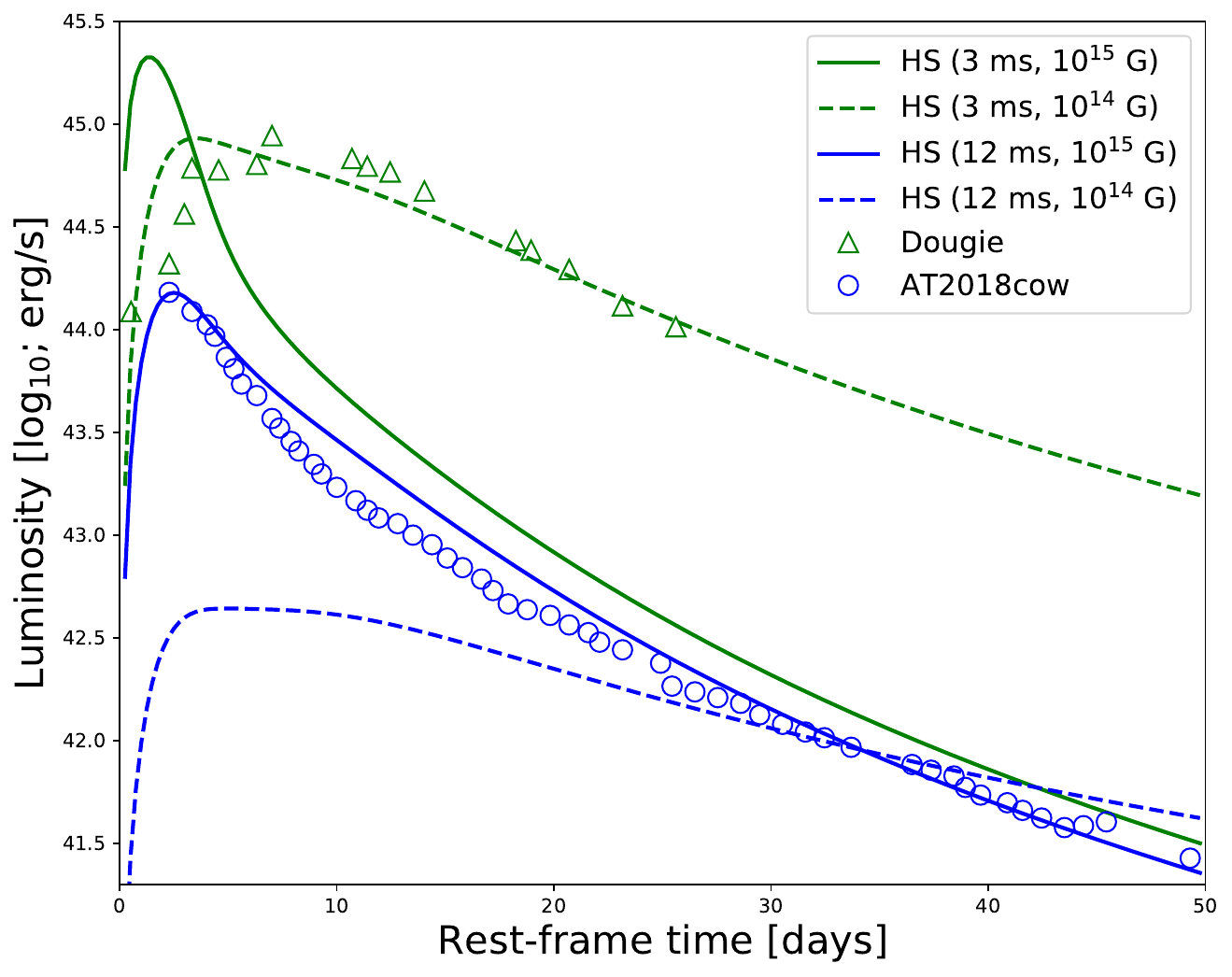}
\includegraphics[scale=0.35]{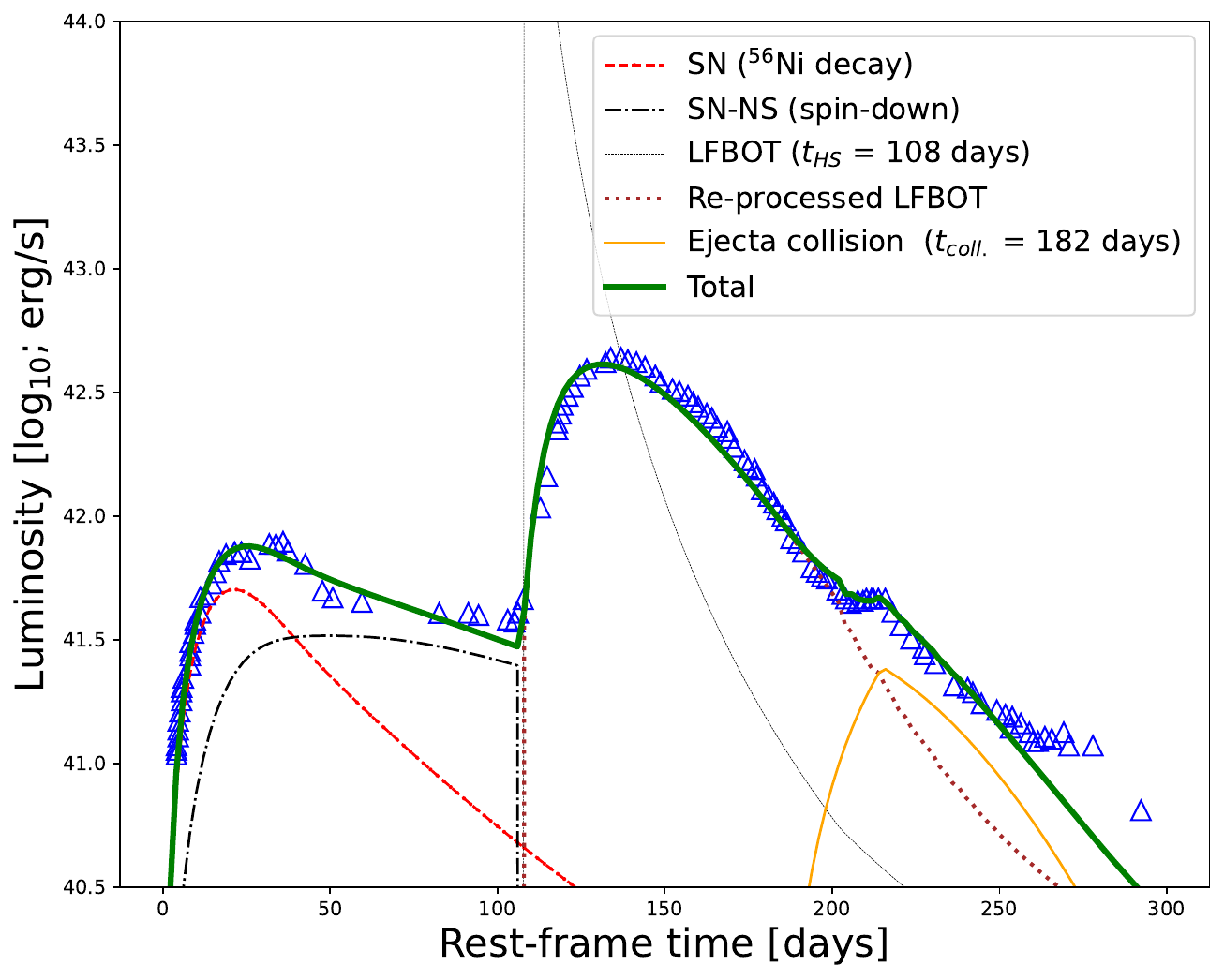}
\caption{{\bf Top panel}: Light curves resulting from the spin-down powering of the ejected outermost layers of the NS ($\sim 0.05M_{\odot}$)  following its conversion to a HS. The models explore different HS spin periods and magnetic field strengths. Shown for comparison are two LFBOTs: AT2018cow and Dougie
 (from \citealt{nicholl_2023}); see \S \ref{sec:HS-spindown} and footnote 1.
 {\bf Bottom panel}: The LFBOT luminosity undergoes re-processing by the expanded SN ejecta, with a subsequent late-time bump resulting from re-heating due to the collision between the QN and SN ejecta.}
\label{fig:LC-LFBOTS}
\end{center}
\end{figure}
%
 %
     
   We now consider the case of no CSM interaction and explore the case of the NS contributing to the plateau via SpD. The fit to the LC is shown
   in Figure \ref{fig:LC-NS-HS} with the corresponding parameters listed in Table \ref{table:table3}. 
   A good fit to the SN peak and the plateau is obtained for any combination of the NS period and magnetic field satisfying the relationship $B_{\rm NS}\sim 7.5\times 10^{13}\ {\rm G}~(P_{\rm NS}/20\ {\rm ms})^2$.    In this case, fits to the second bump can be achieved with a larger mass for the SN ejecta, without requiring an increase in its mass through CSM interaction or plowing, when the opacity 
   is $0.25$  g cm$^{-2}$ during re-brigthening.  Shown in Figure \ref{fig:LC-NS-HS} are two examples of fits with corresponding parameters listed in  Table \ref{table:table3}.  
   A more rigorous implementation of precession (e.g \citealt{zhang_2025} and references therein) is worth exploring in the future.
   
  There is one more important aspect of the HS scenario that must be considered. A partial (only core) conversion of the NS (at $t_{\rm HS}$) ejects on  average $M_{\rm QN, ej}\sim 0.05M_{\odot}$ of 
   the outermost layers. The average kinetic energy of the ejecta, from deposited neutrino
   energy during the partial conversion,  is $E_{\rm QN, KE}\sim 5\times 10^{50}$ ergs  (see \citealt{keranen_2005}
   for details). The corresponding ejecta's speed is $v_{\rm QN, ej.}\sim  0.1c \times (E_{\rm QN, KE}/5\times 10^{50}\ {\rm erg})^{1/2}(0.05M_{\odot}/M_{\rm QN, ej})^{1/2}$. The top panel in Figure \ref{fig:LC-LFBOTS} shows the LC resulting from powering of the QN ejecta by the HS SpD energy which results in rapid transients\footnote{Shown in Figure~\ref{fig:LC-LFBOTS} are two luminous fast blue optical transients (LFBOTs; AT2018cow and Dougie from \citealt{nicholl_2023}) which naturally fit within our model framework.       A NS that experiences the transition to a HS after the SN has dissipated produces a delayed luminous event (an LFBOT) as its SpD energy is deposited into the $\sim 0.05,M_{\odot}$ QN ejecta following the conversion.
Some such events are therefore expected to be observed offset from their birth sites due to natal kicks or dynamical interactions. In extreme cases, the offsets may be large enough for the transient to appear significantly displaced from its host galaxy -- particularly if the host is a low-mass or dwarf galaxy with a shallow gravitational potential.
    This connection we propose between LFBOTs and isolated HS formation could be validated via spectral analysis of data. LFBOTs should eventually exhibit spectra that reflect the composition of the  ejected outermost layers of the NS or of the r-process elements synthesized during its expansion (\citealt{jaikumar_2007,ouyed_2009,ouyed_2011}). Ideally, the model should yield a featureless, blue, blackbody-like spectrum at early times, as observed in LFBOTs  (e.g. \citealt{pursiainen_2018}).}.  We use a QN ejecta's opacity of $0.2$ g cm$^{-2}$ and
    a hard emission leakage factor of $3\times 10^{12}$ s$^{2}$.

    The implication is that the HS SpD power will first be absorbed by the $\sim 0.05M_{\odot}$
  ejecta (re-processed) before re-energizing the expanded SN ejecta.  When $E_{\rm NS, SpD}> E_{\rm SN}$, the supernova ejecta is blown out by the time the transition occurs, thereby unveiling the LFBOT. Otherwise, the LFBOT is buried deep inside the SN ejecta and 
   re-energizes the expanded SN ejecta by depositing energy 
  ($E_{\rm HS, SpD}\sim 2\times 10^{49}\times (30\ {\rm ms}/P_{\rm HS})^{-2}$ erg) into it over a timescale given by the width of the transient ($\sim$ tens of days; see top panel in Figure \ref{fig:LC-LFBOTS}).     Thus, we recover scenario (i) without requiring the HS to collapse into a black hole or relying on excessive gamma-ray leakage when fitting the second peak in the LC.  
    
    This is illustrated in the bottom panel in Figure \ref{fig:LC-LFBOTS} based on the $18M_{\odot}$ SN parameters used for the
    bottom panel in Figure \ref{fig:LC-NS-HS} with the same time delay $t_{\rm HS}=108$ days. The corresponding LFBOT parameters are listed in Table \ref{table:table4}
    prior to re-processing by the expanded SN ejecta. A fit to the second bump is obtained by using the LFBOT luminosity as an input to Eq. (3) in \citet{chatzopoulos_2012} with the opacity of the expanded SN ejecta set to $0.2$ cm$^2$ g$^{-1}$. 
    
    Interestingly, the QN ejecta will eventually catch up with the expanded SN ejecta and shocking it at around $t_{\rm coll.}\sim v_{\rm QN, ej}/(v_{\rm QN, ej}-v_{\rm ej})\times t_{\rm HS}$ 
    after the SN explosion,  causing the late bumps without invoking a precessing HS.  It is calculated using the approach described in
     \S \ref{sec:QN-shock} above with $E_{\rm sh.}= 4\times 10^{-3}E_{\rm QN, KE}$ erg and $t_{\rm sh.}= 33$ days.

         The two SpD sources model offers a compelling framework for interpreting the observed data. The current fits suggest NS magnetic field strengths that exceed the quantum electrodynamics  critical value. To test the robustness of this result, a detailed parameter survey must be conducted to explore whether the observed plateau can also be reproduced with more conventional magnetic field strengths. For now, it suggests two possible channels for magnetar formation: (i) through the classical convection dynamo mechanism (\citealt{duncan_1992}), and (ii) via quark deconfinement, which would generate an even stronger magnetic field from the quark phase, capable of sustaining fields on the order of 
$\sim 10^{18}$ G (see \S \ref{sec:discussion}).

%
 %
\begin{table*}[t!]
\centering
\caption{TARDIS Parameter Settings}
\begin{minipage}{\textwidth}
\begin{tabular}{|c|c|c|c|c|c|c|c|c|c|c|c|c|c|c|c|}\hline
\multicolumn{16}{|c|}{Composition (mass fraction)}  \\\hline
H & He & C & N & O & Ne & Na & Mg & Al & Si & S & Ar & Ca & Cr & Fe & $^{44}$Ti\\\hline
0.010& 0.5153& 0.048& 0.004& 0.250& 0.030& 0.0003 & 0.020& 0.002& 0.03& 0.016& 0.002& 0.002& 0.0004& 0.065& 0.005\\\hline
\end{tabular}
\hfill
\centering
 \begin{tabular}{|c|c|c|c|}\hline
Fit days & Peak 1 (34 days) & Peak 2 (117 days)  & Peak 2 (136 days) \\\hline
\multicolumn{4}{|c|}{SN ejecta parameters }  \\\hline
Luminosity requested (visible range) & $1.3\times 10^{42}$ erg s$^{-1}$ & $4.0\times 10^{42}$ erg s$^{-1}$ & $8.0\times 10^{42}$ erg s$^{-1}$\\\hline
Velocity at the edge of the photosphere & 5000 km s$^{-1}$ & 2300 km s$^{-1}$ & 2000 km s$^{-1}$\\\hline
Velocity at the edge of the ejecta & 9000 km s$^{-1}$  & 7500 km s$^{-1}$ & 7000 km s$^{-1}$\\\hline 
Density ({\it uniform})&  $3.0\times 10^{-14}$ g cm$^{-3}$ &  $8.5\times 10^{-16}$ g cm$^{-3}$\ &  $5\times 10^{-16}$ g cm$^{-3}$\\\hline
\multicolumn{4}{|c|}{Plasma parameters}\\\hline
Ionization mode & lte & nebular\ & nebular\\\hline
Excitation mode & lte & lte & dilute-lte \\\hline
\end{tabular}
 \end{minipage}
\end{table*}
%
 %
\begin{figure}[t!]
\begin{center}
\includegraphics[scale=0.38]{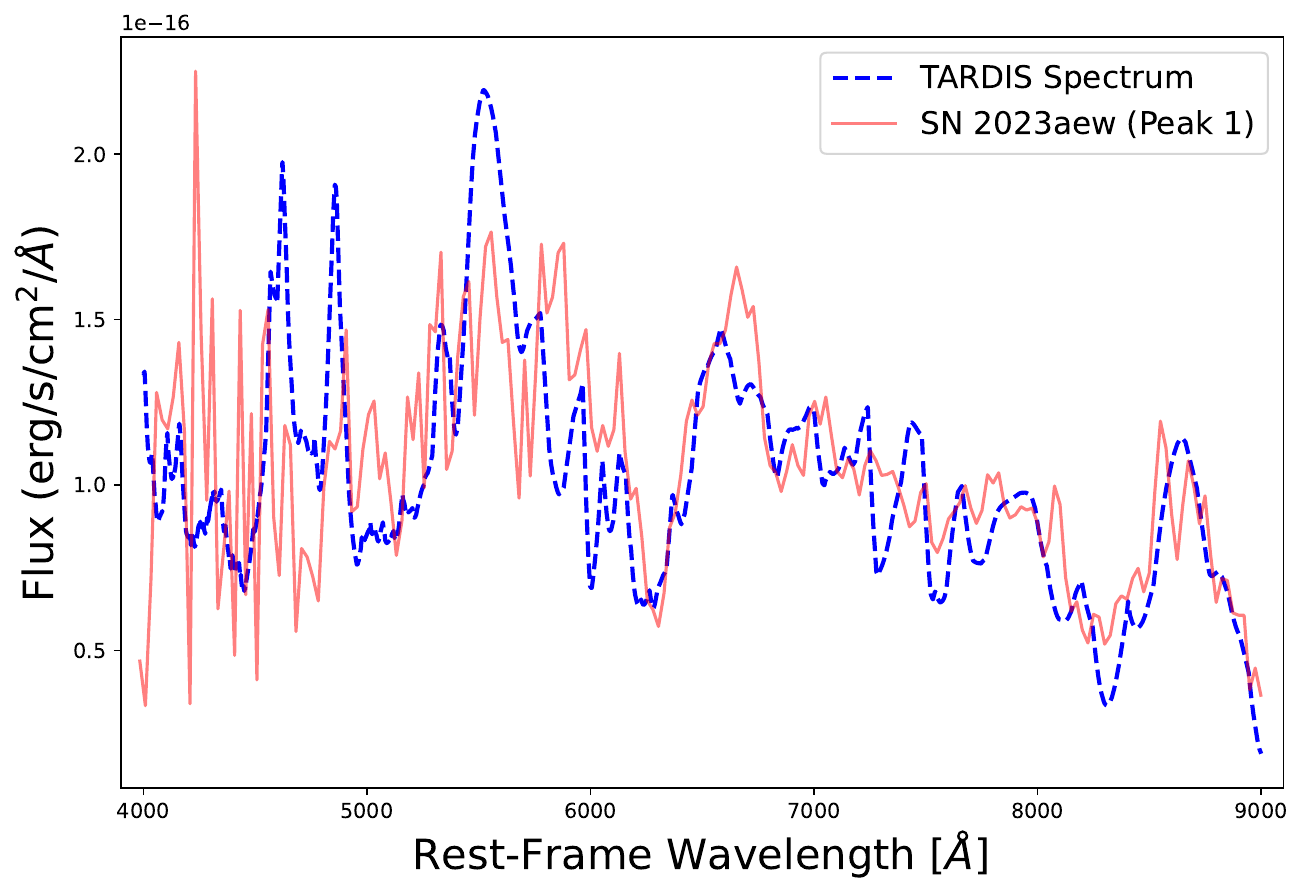}
\includegraphics[scale=0.38]{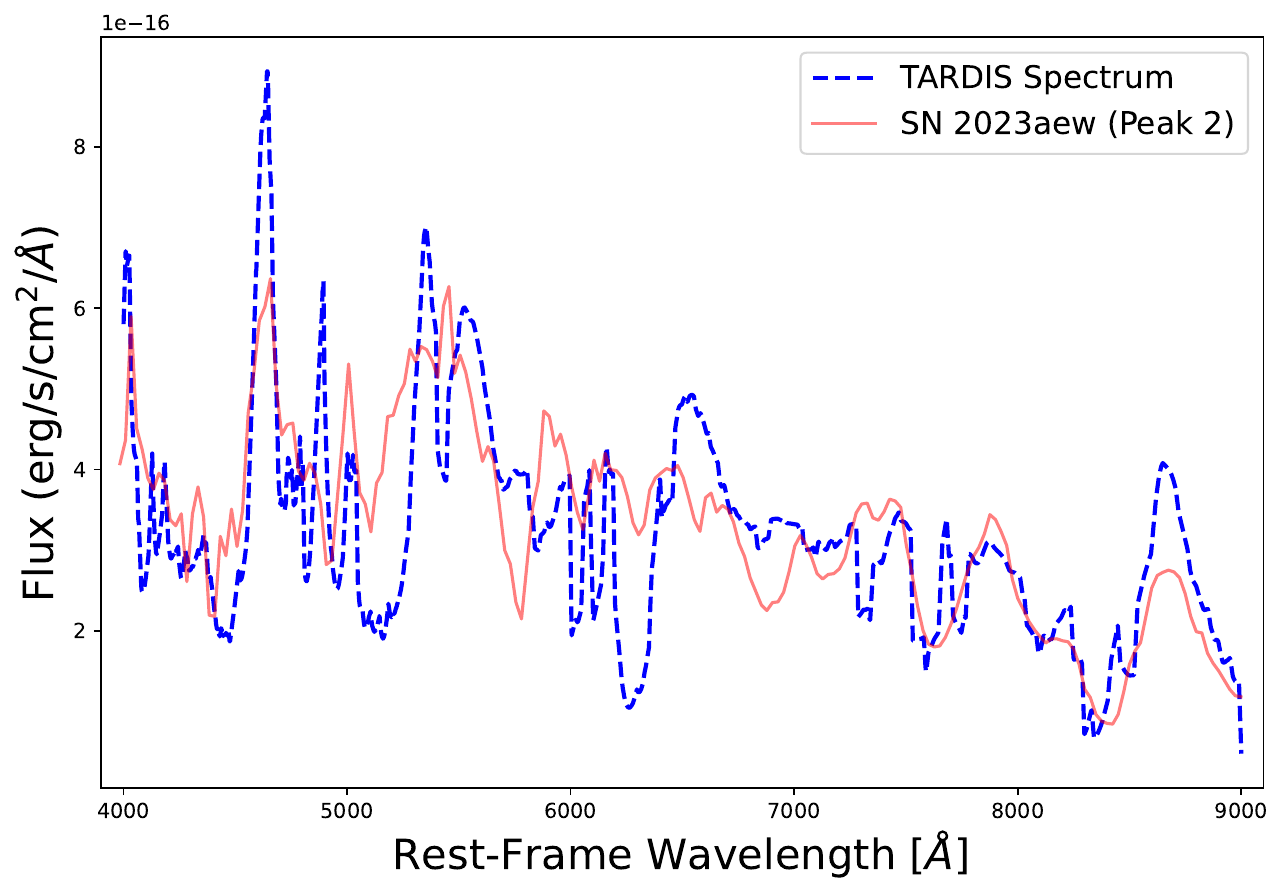}
\includegraphics[scale=0.38]{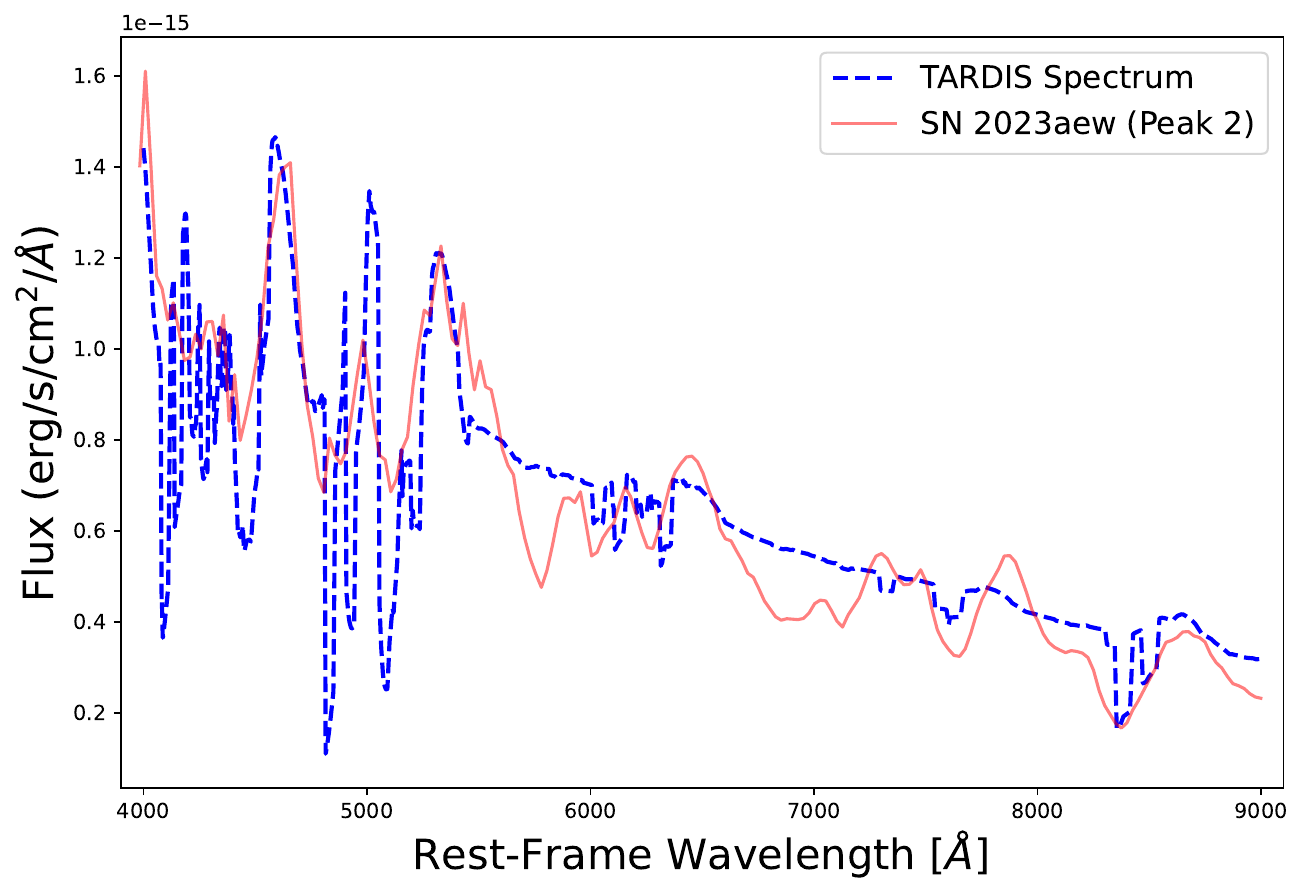}
\caption{Comparison of the observed spectrum of SN 2023aew with the simulated spectrum (see  \S \ref{sec:spectrum}). {\bf Top panel}: at the first peak (34 days after explosion). {\bf Middle panel}: at the second peak (117 days after explosion). {\bf Bottom panel}: at the second peak (136 days after explosion).}
\label{fig:Spectra-Figures}
\end{center}
\end{figure}
%
 %

 ~\\ 
\section{The spectrum}
\label{sec:spectrum}

We utilize a modified version of the Monte Carlo (MC) radiative-transfer code TARDIS  (originally developed for SNe Ia; \citealt{kerzendorf_2014}) repurposed for analyzing Type II SNe (\citealt{vogl_2019} ). It is a one-dimensional, time-independent, MC radiative transfer spectral synthesis code. The code assumes an inner region of completely opaque inner ejecta (i.e. an inner blackbody-emitting photosphere), beneath which all energy injection into the system is assumed to be thermalised. This inner boundary is surrounded by a number of optically thin shells that represent the line-forming region of the ejecta (the computational domain of the simulation). A (user defined) number of photon quanta (r-packets) are created at the inner boundary, and are assigned properties randomly sampled from a single-temperature blackbody. The trajectory of these r-packets are simulated as they traverse the computational domain, with any interaction (either free e-scattering, or bound-bound transitions) simulated. The r-packets that escape the outer boundary of the simulation are used to generate a synthetic spectrum.

User-defined input parameters include specifying: the ejecta's composition, density profile, velocity structure, bolometric luminosity and time since explosion. 
The ejecta's velocity is  set at the edge of the photosphere (inner boundary) and the edge (outer boundary) of the ejecta (the computational domain). While TARDIS is a time-independent code, it is possible to evolve the input parameters to obtain a sequence of self-consistent models, as we do here when fitting SN 2023aew spectra   at around peak 1 (34 days since explosion) and at two times around peak 2 (during the rise time, 117 days since explosion) and at 136 days which is close to the peak. The Spectra of SN 2023aew, based on the Spectral Energy Distribution Machine (\citealt{blagorodnova_2018}),  we obtained from the Weizmann Interactive Supernova Data Repository (\citealt{yaron_2012}). We restrict the analysis to a simple setup adopting a uniform density for the ejecta profile $\propto M_{\rm ej}/(vt)^3$ (which is a good averaging of   the $n=10$ and $\delta=0$  density profile used to fit the  LC).
This gives an average density $\sim 10^{-14}$-$10^{-13}\ {\rm g\ cm}^{-3}$ at  around peak 1 
 and $\sim 10^{-16}$-$10^{-15}\ {\rm g\ cm}^{-3}$ at around peak 2.
 
The fits to the LC suggests a progenitor with ZAMS mass in the $\sim$ 15-16$M_{\odot}$ range  and a SN kinetic energy of a few times $10^{51}$ erg,
with a few solar mass ejecta.  Based on this, for the composition, we adopt that of Type IIb SNe resulting from non-rotating single-star progenitors with masses in the range of 13-21$M_{\odot}$ 
 and explosion energies in the few $10^{51}$  erg range (see \citealt{ergon_2024}).  The composition used in TARDIS (see Table 2) we obtain by mixing the layered composition and averaging it over the progenitors. Elements with lower abundances (i.e., mass fraction  $< 10^{-5}$, such as V, Cr, Mn) were excluded from the fits. The iron mass fraction shown includes Ni and Co. Interestingly, reasonable fits were achieved when the $^{44}$Ti abundance was increased by a factor of 10.

Since our goal is not to perform a quantitative spectroscopic analysis, we have not fine-tuned the model to perfectly match the observations.
For example, the spectra at 34 days after the explosion (i.e. post-peak) is noisy at $\lambda < 5000\AA$  and it is at the limit of TARDIS model conditions which  assumes an inner photosphere where energy packets are emitted from a blackbody surface. We varied only the ejecta and two of the plasma parameters: {\it Ionization} and {\it Excitation} state.  Other plasma parameters such as 
{\it Radiative Rates} and {\it Line Interaction} are set to {\it detailed} and {\it macroatom}, respectively. I.e. 
 radiative rates are calculated using the actual radiation field (as estimated by MC packets during the simulation). The macroatom allows 
detailed modelling of line interactions, including both scattering and fluorescence processes. 
The MC parameters are kept fix to:  {\it Number of Iterations}: 15; {\it Number of Packets}:  $10^5$; {\it Last Number of Packets}: $10^6$; {\it Number of virtual Packets}: 5.   To further simplify the setup and reduce running time, we use only two shells which turned out to be enough to capture key features of the observed spectra.  The runs use the complete Kurucz GFALL atomic dataset (\citealt{kurucz_1995}). The spectrum was generated using the Formal Integral Method (which eliminates nearly all MC noise; \citealt{vogl_2019} and references therein) in the range  $4000 \AA \le \lambda \le 9000 \AA$. 

For the plasma parameters, we set the full Local Thermodynamic Equilibrium (LTE) for excitation and ionization for the 34 day spectrum;
this is the  {\it lte} mode in TARDIS. For peak 2, we use the {\it nebular} mode to solve for ionization populations.  The nebular approximation 
assumes that at $t_{\rm QN}\sim 105$ days after the explosion, the expanded SN ejecta  is optically thin in all ionization continua.  
 In our case, the ejecta becomes optically thin to radiation at time $t_{\rm ej, thin}\sim 1\ {\rm yr}\times (M_{\rm ej}/10M_{\odot})^{1/2} \kappa_{\rm ej, 0.3}^{1/2} v_{\rm ej, 9}^{-1}$
 where $v_{\rm ej, 9}=v_{\rm ej}/10^9\ {\rm cm\ s}^{-1}$ and $\kappa_{\rm ej}$ the opacity in units of 0.3 g cm$^{-2}$. I.e.  
   $t_{\rm QN}\sim t_{\rm thin}$ and the He layer is probably thin enough for the second peak to appear He-poor.
For excitation populations, best fits to peak 2 at 117 days are obtained when we consider that the ejecta is still in an 
  {\it lte} mode. On the other hand, and at 136 days since explosion, reasonable fits require the excitation to be set to {\it dilute-lte}.
 The {\it dilute-lte} prescription in TARDIS acts as an approximation for non-LTE excitation levels again in the optically thin limit. It scales the excitation by a dilution factor to better approximate non-LTE conditions without full treatment.  
  
The resulting fits are shown in Figure \ref{fig:Spectra-Figures}  with the corresponding parameters listed in Table 3. The lower velocities at around peak 2 take into account the slowing down of the SN ejecta after interacting with the CSM.  Despite the many simplifications, the modelled spectra capture key features and the overall trend of the observed spectra in SN 2023aew.
 The P-Cygni  profile at $6000\AA < \lambda < 7000\AA$ in the synthetic spectra at 34 days
  is due to Si II and C II and could be mis-interpreted as due to H$\alpha$ at $6562\AA$ which is used to  suggest the Type IIb
 classification of SN 2023aew. Re-running the spectral analysis without the hydrogen in the composition yields the same final spectra
  implying that the SESN (i.e. first peak) is instead a Type Ib SN.  The exact nature of the first peak observed in SN 2023aew remains uncertain. 
 As noted by \citet{kangas_2024}, the strength of the  $\mathrm{He\,\text{\small I}}$  $\lambda5876+\mathrm{Na\,\text{\small I}}$  $\lambda\lambda5890,5986$ 
 features may be interpreted as indicative of a SN Ic progenitor.   Our synthesized spectrum remains relatively unchanged as the helium content in the ejecta is reduced, though some helium is still required to achieve the best fits -- suggesting a possible Type Ib/Ic classification instead.

  Despite producing a decent fit to the observed spectrum at 117 days, 
  the P-Cygni  profile at $6000\AA < \lambda < 7000\AA$, although reduced, persists  while it is clearly absent in the observed spectrum.
   It  disappears at the 136 days synthesized spectrum when using the {\it dilute-lte} mode for the plasma (see bottom panel in Figure  \ref{fig:Spectra-Figures}).
 This may be understood as due to a decrease in the ejecta's density and opacity.
 One explanation is that by 136 days, parts of the ejecta are still in the {\it nebular} excitation state while some of it is in the {\it dilute-lte} state.
  There is no option to capture such a complex state in TARDIS 
 nor can we include the SN-CSM interaction or any asymmetry in the ejecta when modelling the spectrum. 
  The compositions we used  assume a homologous expansion with a spherically symmetric distribution of the matter
    in thermal and statistical equilibrium (see \citealt{ergon_2024}). 
  Considering a non-uniform density profile and composition of the ejecta may also help address some of the limitations of the approach used here. Additionally, as noted earlier, the neutron-rich QN ejecta is favorable for the nucleosynthesis of heavy elements (see \citealt{jaikumar_2007}) . If not fully dissociated upon impact with the SN ejecta, this material could introduce spectral signatures that are not accounted for in the present analysis.

 The fits required  a $^{44}$Ti  mass fraction higher than what was given by the 1D models (by a factor of 10; see Table 2). This is justifiable 
  given the underproduction of $^{44}$Ti in 1D models as compared to measured values
  and yields from multidimensional SN models (see e.g. \citealt{the_2006,sieverding_2023} and references therein).
  This may also be consistent with the fit to the first peak in SN 2023aew LC which suggests an energetic explosion where 
  ejection of more matter during the SN, including $^{44}$Ti, is to be expected due to a shift in the mass cut;  
 although it  is also sensitive  to other parameters such as the metallicity (e.g. \citealt{woosley_1995}).

\section{Discussion and conclusion}
\label{sec:discussion}

While the following discussion focuses on delayed energy input from the conversion of a NS to either a QS or a HS, any alternative mechanism capable of depositing $\sim 2\times 10^{49}$~erg of heat over a period of $\sim 40$~days into the extended envelope should, in principle, produce similar results.
The conversion timescale, $t_{\rm QN}$ or $t_{\rm HS}$, we suggest is linked to quantum nucleation of quark centers in the 
NS core. I.e., nucleation driven by  quantum tunnelling effects which is dominant in cold ($T < 1$ MeV) and deleptonized NSs which we consider here (e.g. \citealt{horvath_1992,olesen_1993,heiselberg_1993,iida_1997,harko_2004,bombaci_2004}). 
  To convert the core of the NS  into quark matter,  at least one droplet of quark matter (of critical size) must be formed. 
The nucleation rate per unit volume  (e.g. \citealt{olesen_1993} and references therein) is $\Gamma \sim \mu_q^4 \exp\left(-\frac{16 \pi \sigma^3}{3 (\Delta P)^2 \mu_q}\right)$
    where $\mu_q$ is the quark chemical potential in MeV (switching to natural units where $\hbar=c=1$).
The nucleation timescale, defined as the time it takes for the nucleation of one single critical bubble,  
      is $\tau \sim 1/(\Gamma V_\text{c})$ where $V_\text{c}=\frac{4\pi}{3}R_{\rm c}^3$ is the volume of the nucleation region (here the $R_{\rm c} \sim 1$ km NS core).     Once nucleation is triggered the growth of a macroscopic quark core  occurs on dynamical timescale $R_{\rm c}/c\sim 10^{-6}$-$10^{-5}$ s.
           
 The nucleation rate is extremely sensitive to the  surface tension of the quark droplet ($\sigma$)  and the pressure difference ($\Delta P$) between deconfined quark matter and hadronic nuclear matter  (i.e.  on the  EOS).  The estimated range of values for the tension is $\sigma \approx$ 10-300  MeV fm$^{-2}$
      (e.g. \citealt{heiselberg_1993,iida_1998,alford_2001,voskresensky_2003}) 
    and $\Delta P \approx$ 1-150 MeV fm$^{-3}$ (e.g. \citealt{weber_2005} and references therein) means that nucleation timescale varies widely.  
   
  For $\mu_{\rm q}=340$ MeV (expected in the core of NSs), using $\Delta P = 2.6 (\sigma_{30}^3/\mu_{340})^{1/2}$ MeV fm$^{-3}$ yields $t_{\rm QN}\sim 132$ days; 
 $\mu_{\rm q}\sim 340.4$ MeV gives $t_{\rm QN}\sim$ 105-109 days as needed for SN 2023aew.  Such a small, but a non-zero value for $\Delta P$, hints at  
   EOS with marginal stability between the hadronic and quark matter. 
   In general, for  $\mu_{\rm q}< 350$ MeV  the nucleation timescale ($t_{\rm QN}$ or $t_{\rm HS}$ in our model)
 is less than one year. Therefore, if our model is correct, the time delay between the two peaks in SN 2023aew-like SNe could potentially be used to constrain key properties of quark matter, such as the surface tension ($\sigma$) and the EOS for both hadronic and quark matter.
  
If the conversion is via the QN shock, the energy input is from the latent heat when the entire NS converts to a stable QS (e.g. \citealt{itoh_1970,bodmer_1971,terazawa_1979,witten_1984})
yielding $\sim 10^{52}$ erg in QN kinetic energy (see \citealt{ouyed_2022a,ouyed_2022b} and references therein). 
The expanding relativistic QN ejecta, with mass $M_{\rm QN, ej} \sim 10^{-5}M_{\odot}$, quickly becomes optically thin and ceases to expand significantly. This occurs well before it catches up with the preceding SN ejecta, at which point the QN ejecta's rest-frame number density is 
 $n^\prime_{\rm QN, ej}\sim 10^{14}\ {\rm cm}^{-3}$  (see \citealt{ouyed_2009} and Appendix B in \citealt{ouyed_2020} for details).

Conversion of the QN kinetic energy requires an efficient reverse shock (RS) when the QN and the SN ejecta collide. 
This is satisfied if $n^\prime_{\rm QN, ej}/n_{\rm ej} < \Gamma_{\rm QN}^2$ (e.g. \citealt{landau_1959,meszaros_1992})
where $n_{\rm ej}$ is the SN ejecta's number density at $t_{\rm QN}$ and $\Gamma_{\rm QN}\sim 10^{3}$ the Lorentz factor
of the QN ejecta. Thus heating of the expanded SN ejected occurs when $n_{\rm ej} > 10^8$ cm$^{-3}$ which is equivalent to  $t_{\rm QN} <  t_{\rm QN, RS}\sim 1\ {\rm yr}\times (M_{\rm ej}/10M_{\odot})^{1/3}/v_{\rm ej, 9}$. For $t_{\rm QN} > t_{\rm QN, RS}$,
the forward shock (FS) dominates, leading to a Gamma-Ray Burst (GRB) in the QN model (see \citealt{ouyed_2020}).\footnote{In the QN model, GRBs occur when the NS blows out  the SN ejecta \emph{before}  it converts to a QS or a HS. This scenario arises when the NS SpD energy exceeds the SN energy, $E_{\rm NS, SpD} > E_{\rm SN}$; i.e. $P_{\rm NS} < \sqrt{2}/E_{\rm SN ,52}^{1/2}$ ms. It involves a rapidly rotating NS which results in a broad Type Ic SN. The interaction between the QN ejecta -- or the HS wind -- and the blown-out turbulent, low-density SN ejecta produces a GRB and a Band spectrum, as described in \citet{ouyed_2020}; see Figure 6 in that paper. Conversely, when $E_{\rm NS, SpD} < E_{\rm SN}$, as is the case for a slowly rotating NS, the NS contributes energy to the SN (e.g. producing the plateau in the LC) without blowing out the SN ejecta and yielding a lower-velocity ejecta compared to the $E_{\rm NS, SpD} > E_{\rm SN}$ case. The delayed energy input -- from either the QN shock or HS SpD -- occurs within a denser SN ejecta, ultimately giving rise to a superluminous second bump as in SN 2023aew.}
For significantly longer delays -- after the SN ejecta has fully dissipated -- the QN model instead predicts the production of Fast Radio Bursts (FRBs) (\citealt{ouyed_2025}). 

If the neutron-to-quark transition proceeds as a smooth crossover, no latent heat is released, favoring the HS scenario. In this case, the HS must possess a magnetar-strength surface magnetic field. One possibility is the delayed emergence of a buried magnetic field resulting from fallback accretion during the supernova explosion. However, such a scenario would require an unrealistically large amount of fallback material (e.g. \citealt{torres_2016}), and the associated magnetic field emergence timescale (\citealt{geppert_1999}) would far exceed the observed time delays between the peaks in these SNe. An alternative explanation involves the formation of a quark phase in the HS core capable of sustaining magnetic fields on the order of $\sim 10^{18}$ G. Based on simple energy scale arguments, quark matter at energy scales of $\sim$ 250 MeV can support such extreme magnetic fields, compared to the $\sim 10^{15}$ G 
 limit for hadronic matter, which is associated with energy scales of $\sim$ 8 MeV.  In particular, a quark phase exhibiting color ferromagnetism, where a chromomagnetic field spontaneously forms during the phase transition, could generate such fields (see \citealt{iwazaki_2005,dvornikov_2016}). This phenomenon is analogous to conventional ferromagnetism in condensed matter physics -- where electron spins align in response to a magnetic field -- but here involves the alignment of quark color magnetic moments in response to a chromomagnetic field. Beyond its implications for QCD, such a phase provides a novel channel for the delayed formation of magnetars (i.e., HSs in our model), with potentially significant consequences for astrophysical observations.

If the interpretation we presented here is correct, it strengthens the case for isolated NSs converting to HSs (yielding LFBOTs) as a viable $r$-process site in the universe \citep{keranen_2005,jaikumar_2007,ouyed_2009,ouyed_2011}. Intriguingly, some double-peaked SNe and SLSNe may harbor unresolved LFBOT-like components, suggesting a broader population of such events awaiting discovery. In our model, the energy associated with LFBOTs should be comparable to that of double-peaked SLSNe, since the latter are assumed to result from the reprocessing of LFBOT luminosity by the expanded SN ejecta. The presence of coherent radio emission mechanisms in these highly magnetized HSs could also provide a natural link to FRBs, as discussed in \citet{ouyed_2025} for the case of a full NS-to-QS conversion. This connection can be directly tested by searching for FRB-like signals coincident with LFBOTs, offering a novel observational probe into their central engines.

In addition to these electromagnetic signatures, multimessenger signals such as gravitational waves and neutrinos could offer further constraints on the nature of the NS-to-HS (or QS) transition. For example, if the conversion involves a rapid core reconfiguration, it may produce a faint but detectable gravitational wave burst, particularly in nearby events. Likewise, if deconfinement occurs under conditions of neutrino trapping, a brief neutrino flash may accompany the transition (see \citealt{ouyed_2022a,ouyed_2022b} and references therein). While challenging to detect, such signals would provide independent confirmation of the underlying engine and help distinguish between explosive and non-explosive conversion scenarios. Together, electromagnetic and multimessenger observations could thus provide a powerful test of our model and offer new insight into compact object evolution and transient phenomena.

\section*{Acknowledgements}
This work is dedicated to the memory of Prof. Jan Erling Staff, a colleague, former student, and friend who is dearly missed.

\section*{DATA AVAILABILITY}

No new data were generated or analysed in support of this research.

\end{document}